\documentclass[12pt]{article} 
\usepackage{latexsym,amsfonts}
\newcommand{\rr}{{\mathbb{R}}}
\newcommand{\cc}{{\mathbb{C}}}
\newcommand{\sll}{SL(2,$\rr$)}
\newcommand{\slu}{SL(2,$\rr$)/U(1)}
\renewcommand{\d}{{\rm d}}
\newcommand{\e}{{\rm e}}
\renewcommand{\i}{{\rm i}}
\newcommand{\delz}{\partial_z}
\newcommand{\zb}{{\bar{z}}}
\newcommand{\delzb}{\partial_\zb}
\newcommand{\delzp}{\partial_{z'}}
\newcommand{\dels}{\partial_\sigma}
\newcommand{\delsp}{\partial_{\sigma'}}
\newcommand{\thr}[1]{\tanh^{#1}\!r\;}
\newcommand{\sh}[2]{\sinh^{#1}\!{#2}\;}
\newcommand{\shr}[1]{\sh{#1}{r}}
\newcommand{\ch}[2]{\cosh^{#1}\!{#2}\;}
\newcommand{\chr}[1]{\ch{#1}{r}}
\newcommand{\Ab}{\bar{A}}
\newcommand{\Bb}{\bar{B}}
\newcommand{\tr}{{\rm tr}}
\newcommand{\intl}{\int\limits}

\begin{document}

\title{\makebox[0cm]{
\parbox{\textwidth}{\centering
Analytical Solution of the \slu{} WZNW Black Hole Model}}
\makebox[0cm]{\raisebox{3cm}[0cm][0cm]
{\parbox{\textwidth}
{\normalsize\noindent DESY 96--221\hfill ISSN 0418-9833\\
December 1996\hfill hep-th/9702095
}}}
}
\author{Uwe M\"uller\thanks{E-mail: {\tt
      umueller@ifh.de}}\hspace{0.5ex} and\setcounter{footnote}{6}
  Gerhard Weigt\thanks{E-mail: {\tt weigt@ifh.de}}\\ Deutsches
  Elektronensynchrotron DESY\\ Institut f\"ur Hochenergiephysik IfH
  Zeuthen\\ Platanenallee 6, D--15738 Zeuthen, Germany}

\date{}
\maketitle

\begin{abstract}
  
  The gauged \sll/U(1) Wess-Zumino-Novikov-Witten (WZNW) model 
  is classically an integrable conformal field theory. 
  We have found a Lax pair representation for the non-linear 
  equations of motion, and a B{\"a}cklund transformation. 
  A second-order differential equation of the Gelfand-Dikii type 
  defines the Poisson bracket structure of the theory, and its 
  fundamental solutions describe the general solution of
  the WZNW model as well. 
  The physical and free fields are related by non-local 
  transformations. The (anti-)chiral component of the 
  energy-momentum tensor which factorizes into conserved 
  quantities satisfying a non-linear algebra characteristic for 
  parafermions assumes the canonical free field form. 
  So the black hole model is expected to be prepared for an exact 
  canonical quantization. 
\end{abstract}

Witten has shown \cite{Witten} that the conformal \slu{} gauged 
WZNW action \cite{BA} describes string propagation in the 
background of a two-dimensional space-time black hole. He 
conjectured that ``the conformal field theory governing the 
black hole is more or less exactly soluble.'' However, this 
theory is not conformal in the 
ordinary sense. It is true that it has classically a traceless 
energy-momentum tensor, but the $\beta$-function is non-zero 
\cite{Callan,Witten,DVV,Tseytlin} as the gauged WZNW action 
corresponds to a $\sigma$-model with curvature. The inherent 
non-local structure of the theory might settle this question. 
But that requires, obviously, a complete analytical solution of 
the theory, and its exact quantization. Classical solutions were found
for two-dimensional string theory in any curved space-time \cite{BS}.
But we are looking here for unconstrained field-theoretic solutions.
Our intention is to find a canonical transformation of the physical
fields onto free fields as a prerequisite for a canonical quantization
of the present non-linear theory. 

Interestingly, there is an embedding of the black hole in an 
integrable non-abelian Toda theory \cite{GS}. Although this apparently
does not solve the addressed problem the results of refs.\ \cite{GS} 
and \cite{Bilal} stimulated the analytical calculations presented in 
this letter. We could show that the analytical solution of the \slu{} 
coset theory is 
asymptotically related to the solution of that non-abelian Toda theory
\cite{buckow}. However, there is an independent way of proving 
integrability of the \slu{} model. We have found a Lax pair 
representation for the non-linear equations of motion, and a 
B{\"a}cklund transformation as well. We have solved the equations in 
general, and derived a non-local transformation of the 
interacting fields onto free fields for periodic boundary conditions. 

In this letter we discuss the classical theory. Here we will not 
be able to give all the details and 
proofs, in particular, we shall omit the Lax pair representation 
completely and refer to a separate paper \cite{prep}.

The \slu{} gauged WZNW action can be derived from the U(1) gauge
invariant WZNW action. In light-cone coordinates $z=\tau+\sigma$, 
$\zb=\tau-\sigma$, $\delz=(\partial_\tau+\partial_\sigma)/2$,
$\delzb=(\partial_\tau-\partial_\sigma)/2$ this action is
($\gamma\equiv\sqrt{2\pi/k}\,$)
\begin{eqnarray}
S[g,A]&=&\frac{1}{\gamma^2}\intl_M\Bigg[
\frac{1}{2}\tr\left(g^{-1}\delz g g^{-1}\delzb g\right)+
\tr\left(Ig^{-1}\delz g\right)A_\zb+\nonumber\\
&&\hspace{1cm}+\;
\tr\left(I\delzb gg^{-1}\right)A_z+
\left(\tr\left(IgIg^{-1}\right)-2\right)A_zA_\zb\Bigg]
\d z\d\zb+\nonumber\\
&&+\;\frac{1}{6\gamma^2}\hspace{-2.8ex}\intl_{B,\;\partial B=M}\hspace{-2.8ex}
\tr\left(g^{-1}\d g\wedge g^{-1}\d g\wedge g^{-1}\d
g\right).
\end{eqnarray}
Here $I$ is a matrix
\begin{equation}
I=\left(\begin{array}{rr}0&1\\\hspace{-1ex}-1&0\end{array}\right),
\end{equation}
and the Lie group-valued field $g$ is parameterized by 
($\sigma_3$ is the third Pauli matrix) 
\begin{equation}\label{para}
 g=g(r,t,\alpha)=\exp\left((\alpha+t)I/2\right)
\exp(r\sigma_3)\exp\left((\alpha-t)I/2\right).
\end{equation}

In order to define the gauged WZNW action we have to eliminate the 
abelian gauge field $A_z$, $A_\zb$. This will not be done by path
integration \cite{Buscher,Witten,Tseytlin}. Such integrations are 
still incomplete because functional determinants remain uncalculated
\cite{Buscher}, and we would not get a suitable effective action to 
start with. Instead, we eliminate the gauge field through its 
equation of motion and gauge the angle variable $\alpha$ by 
$\alpha=0$. So we obtain an exact classical action \cite{BA}
\begin{equation} \label{action}
S[r,t]=\frac{1}{\gamma^2}\intl_M
\left(\delz r\delzb r+\thr{2}\delz t\delzb t\right)\d z\d\zb.
\end{equation}
Its (euclidean signature) black hole \cite{Witten} target-space 
metric
\begin{equation}
  \d s^2=\d r^2+\thr{2}\d t^2
\end{equation}
has the form of a semi-infinite cigar which is asymptotic for 
$r\to\infty$ to $\rr\times S^1$ with a flat metric. The conformal 
invariance of this theory is indicated by the tracelessness of its 
energy-momentum tensor.

The equations of motion following from (4) are 
\begin{eqnarray}\label{eom}
  \delz\delzb r&=&\frac{\shr{}}{\chr{3}}\delz t\;\delzb t ,\nonumber\\ 
  \delz\delzb t&=&-\frac{1}{\shr{}\chr{}} \left(\delz r\;\delzb
  t+\delz t\;\delzb r\right).
\end{eqnarray}
They provide conservation of the chiral component of the
energy-momentum tensor (we will not indicate, whenever possible,  
the similar anti-chiral parts)
\begin{equation}\label{emt}
  T=\frac{1}{2}\left(T_{\tau\tau}+T_{\tau\sigma}\right)=
  \frac{1}{\gamma^2}\left((\delz r)^2+\thr{2}(\delz t)^2\right).
\end{equation}

Moreover, there are two further conserved chiral quantities
on shell
\begin{equation}\label{vpm}
  V_{\pm}=\frac{1}{\gamma^2}
  \e^{\pm\i\nu}\left(\delz r\pm\i\thr{}\delz t\right).
\end{equation}
For the conservation law $\delzb V_{\pm}=0$, the (non-local) 
$\nu$ is sufficiently defined by the derivatives \cite{GS}
\begin{equation}\label{nu}
  \delz\nu=(1+\thr{2})\delz t,\quad\delzb\nu=\chr{-2}\delzb t.
\end{equation}
However, it will be easy to integrate them, once we have got the 
general solution of the equations of motion, since the  
integrability conditions of (\ref{nu}) just yield the second equation 
of (\ref{eom}). Surprisingly, the chiral component of the
energy-momentum tensor factorizes in terms of the
$V_{\pm}(z)$ and looks like a Sugawara construction 
\begin{equation}
  T=\gamma^2V_+V_-.
\end{equation}
But the conformal spin-one quantities $V_\pm$ 
are no usual Kac-Moody currents.

The general solution of the equations of motion will be stated 
here without proof \cite {prep}. It can easily be seen that
\begin{eqnarray}\label{solution}
  \shr{2}&=&X\bar{X},\nonumber\\ 
  t&=&\i(B-\Bb)+\frac{\i}{2}\ln\frac{X}{\bar{X}}
\end{eqnarray}
solve the equations of motion (\ref{eom}) if
\begin{eqnarray}\label{tosolution}
  X&=&A+\frac{\Bb'}{\Ab'}(1+A\Ab),\nonumber\\ 
  \bar{X}&=&\Ab+\frac{B'}{A'}(1+A\Ab).
\end{eqnarray}
$A(z)$, $B(z)$, $A'(z)$, $B'(z)$ ($\Ab(\zb)$, $\Bb(\zb)$, $\Ab'(\zb)$,
$\Bb'(\zb)$) are arbitrary chiral (anti-chiral) functions and their
derivatives, respectively. The solution (\ref{solution},
\ref{tosolution}) is invariant under $GL(2,\cc)$ transformations
\begin{eqnarray}\label{moebius}
  A&\to&T[A]=\frac{aA-b}{cA+d} ,\nonumber\\ 
  B&\to&T[B]=B+\ln(c
A+d),\nonumber\\ 
  \Ab&\to&T[\Ab]=\frac{d\Ab-c}{b\Ab+a} ,\nonumber\\ 
  \Bb&\to&T[\Bb]=\Bb+\ln(b\Ab+a), \\ &&\left(
\begin{array}[]{cr}
  a&-b\\ c&d
\end{array}\right)
\in {\rm GL}(2,\cc).\nonumber
\end{eqnarray}

In order to find free field realizations  of these functions, first
we have to calculate their Poisson brackets using the canonical 
Poisson brackets of the physical fields $r$, $t$ (at equal $\tau$!). 
The procedure applied in ref.\ \cite{Bilal} is an indirect one and 
does not prove a unique result. We shall calculate the Poisson 
brackets directly by using the characteristic differential equation
\begin{equation}\label{chardgl}
  y''-(\delz V_-/V_-)y'-\gamma^2Ty=0,
\end{equation}
which is derived from the conserved quantities (\ref{vpm}). (We should 
mention here that this equation fits into the Gelfand-Dikii
hierarchy\footnote{We thank Adel Bilal for discussions about the
  Gelfand-Dikii hierarchy.}
with $W^{(1)}=(\delz V_-/V_-)$ and that it could be transformed with 
a modified energy-mo\-men\-tum tensor $\tilde{T}$ into
\begin{equation}
  \tilde{y}''-\tilde{T}\tilde{y}=0,
\end{equation}
well-known in Liouville theory.
$y$ has conformal weight zero, $\tilde{y}=y/\sqrt{V_-}$ conformal 
weight $-1/2$.)
To get $A(z)$, $B(z)$ related to the solutions of this equation, 
we have to integrate the eqs.\ (\ref{nu}) so that the $V_{\pm}$ become
defined. From the general solution (\ref{solution}, \ref{tosolution}) 
one obtains 
\begin{equation}\label{nuint}
 \nu-t=\i(B+\Bb)-\frac{\i}{2}\ln\frac{1+X\bar{X}}{(1+A\Ab)^2},
 \qquad \nu+\bar{\nu}=2t.
\end{equation}

We have, furthermore, to rewrite the coefficients of
eq.\ (\ref{chardgl}) completely in terms of $A(z)$, $B(z)$ using 
again (\ref{solution}, \ref{tosolution}). The two fundamental 
solutions of eq.\ (\ref{chardgl}) are then given by
\begin{eqnarray}\label{fundsol}
  y_1(z)&=&\e^{B(z)} ,\nonumber\\ 
  y_2(z)&=&y_1(z)\int^{z}\frac{W(z')}{y_1^2(z')}\d z'.
\end{eqnarray}
The Wronski determinant
\begin{equation}\label{Wronski}
  W=y_1y_2'-y_2y_1'
\end{equation}
is here related by means of eq.\ (\ref{chardgl}) to $V_-$
\begin{equation}
  W=\gamma^2V_-=A'\e^{2B},
\end{equation}
so that we find $A(z)$ and $B(z)$  explicitly in terms of the 
fundamental solutions
\begin{equation}\label{Ainy}
  A(z)=\frac{y_2}{y_1}, \qquad B(z)=\ln{y_1}.
\end{equation}
The Poisson brackets of the $A(z)$, $B(z)$ can be 
calculated once we have got those of the ${y_1}$, ${y_2}$, and,
finally, we could read off the $A(z)$ and $B(z)$ as functions 
of free fields.

The differential equation (\ref{chardgl}) determines the dependence 
of the $y_i$ on the fields $r$, $t$ and their momenta ${\pi_r}$, 
${\pi_t}$. In order to derive the Poisson brackets, we need the 
variations $\delta y_i$ in terms of the variations $\delta r$, 
$\delta t$, $\delta {\pi_r}$, and $\delta {\pi_t}$, which enter 
the definition of the Poisson bracket. Varying eq.\ (\ref{chardgl}) 
we obtain
\begin{equation}\label{var-y-eq}
\delta y''-(\partial V_-/V_-)\delta y'-\gamma^2T\delta y=
\delta(\partial V_-/V_-)y'+\gamma^2\delta Ty.
\end{equation}
The homogeneous part of this differential equation is identical to
eq.\ (\ref{chardgl}). To find a unique solution of the inhomogeneous 
equation we need subsidiary conditions. Using field-theoretic initial 
conditions (no zero modes!)
$\delta {y_i}(-\infty)=0$, $\delta {y'_i}(-\infty)=0$, it can be
represented by the two solutions $y_1$ and $y_2$ of the homogeneous
equation 
\begin{eqnarray}\label{var-y-sol}
\delta y_i(\sigma)&=&\int_{-\infty}^\sigma{\rm d}\sigma'
\Omega(\sigma,\sigma')
\left(\delta(\partial V_-/V_-)(\sigma')y_i'(\sigma')+
\gamma^2\delta T(\sigma')y_i(\sigma')\right),\nonumber \\
\Omega(\sigma,\sigma')&\equiv&
\frac{y_1(\sigma')y_2(\sigma)-y_2(\sigma')y_1(\sigma)}
{y_1(\sigma')y_2'(\sigma')-y_2(\sigma')y_1'(\sigma')}.
\end{eqnarray}
This leads directly to the Poisson brackets of the $y_i$ \cite {prep}
\begin{equation}\label{Poisson-fixed}
\{y_i(\sigma),y_j(\tilde\sigma)\}=\frac{\gamma^2}{2}
\left(
y_i(\sigma)y_j(\tilde\sigma)-y_j(\sigma)y_i(\tilde\sigma)\right)
\epsilon(\sigma-\tilde\sigma).
\end{equation}
$\epsilon(\sigma)$ denotes here the sign function. 

The Poisson brackets for the $A(z)$, $B(z)$ are found by taking into 
consideration eqs.\ (\ref{Ainy}), and we read off the free 
field realizations \cite{prep}
\begin{eqnarray}\label{ABinphi}
  A'(z)&=&\gamma\left(\delz \phi_1-\i\delz\phi_2\right)
  \exp\left(-2\gamma\phi_1(z)\right),\nonumber\\
  B(z)&=&\gamma\left(\phi_1-\i\phi_2\right).
\end{eqnarray}
The conserved $V_\pm(z)$ then become 
\begin{equation}\label{Vpminphi}
  V_\pm(z)=\frac{1}{\gamma}\left(\delz\phi_1\pm\i\delz\phi_2\right)
  \exp\left(\pm 2\i\gamma\phi_2(z)\right),
\end{equation}
and, as expected, the energy-momentum tensor $T$ assumes the 
canonical free field form 
\begin{equation}\label{Tinphi}
  T=\left(\delz\phi_1\right)^2+\left(\delz\phi_2\right)^2.
\end{equation}

Vice versa, we could get the same free field realizations 
(\ref{ABinphi}), consistently, as solutions of eq.(\ref{chardgl}) 
if its coefficients would have been given by (\ref {Vpminphi}, 
\ref {Tinphi}), and, third, by requiring that the free fields 
of the energy-momentum tensor (\ref {Tinphi}) are local in $y_i(z)$
and their derivatives \cite{prep}. This remains true for the periodic 
string case, after including the corresponding zero modes. But for 
periodic boundary conditions the non-periodic sign function has to be 
replaced by the periodic sawtooth function
\begin{equation}\label{saw}
h(\sigma)=\epsilon(\sigma)-\frac{1}{\pi}\sigma,
\end{equation}
and we have to take into consideration the periodic $\delta$-function
\begin{equation}\label{delta}
{\delta}_{2\pi}(\sigma)\equiv\frac{1}{2}\dels\epsilon(\sigma)
=\sum_{n}\delta(\sigma-2\pi n).
\end{equation}
$\epsilon(\sigma)$ denotes now the stair-step function
\begin{equation}\label{epsi}
\epsilon(\sigma)=2n+1 \quad \mbox{for} \quad 2n\pi<\sigma<(2n+1)\pi,
\end{equation}
which coincides with $\rm sign(\sigma)$ for $-2\pi<\sigma<2\pi$.
(The principal value is understood at the discontinuities.)

It is worth mentioning here that the conserved $V_\pm(z)$ are related
to the \slu{} coset currents. They satisfy non-linear Poisson brackets
\begin{eqnarray}\label{valgebra}
  \{V_\pm(\sigma),V_\pm(\sigma')\}&=&\gamma^2
  V_\pm(\sigma)V_\pm(\sigma')\epsilon(\sigma-\sigma'),\nonumber\\ 
  \{V_\pm(\sigma),V_\mp(\sigma')\}&=&-\gamma^2
  V_\pm(\sigma)V_\mp(\sigma')\epsilon(\sigma-\sigma')+
  \frac{1}{\gamma^2}\delta'(\sigma-\sigma')
\end{eqnarray}
which are characteristic for parafermions \cite{FZ,BA}.\footnote{We
thank Konstadinos Sfetsos for a comment on parafermions
in this respect. Instead of solving the present conformal theory by
an infinite set of conservation laws, we are intending to emphasize
the dynamics of the theory governed by the general solutions of its
equations of motion.} 
This non-linear algebra also provides the Virasoro algebra, and
implies conformal weight one for the $V_\pm$
\begin{equation}
  \{T(\sigma),V_\pm(\sigma')\}=-\left( \delsp
  V_\pm(\sigma')\delta(\sigma-\sigma')-
  V_\pm(\sigma')\delta'(\sigma-\sigma')\right).
\end{equation}

It remains to calculate $A(z)$ by integrating $A'(z)$. We choose now
(closed string) periodic boundary conditions, and in the target space
\begin{eqnarray}\label{bc}
  r(\tau,\sigma+2\pi)&=&r(\tau,\sigma), \nonumber\\ 
  t(\tau,\sigma+2\pi)&=&t(\tau,\sigma)+2\pi w,\quad w\in{\mathbb{Z}}.
\end{eqnarray}
As $t(z)$ is an angle variable in the parameterization (\ref{para}),
$w$ is a `winding number' describing how often the string surrounds the
origin of the target space.  
In the following we restrict our general solution to
real free fields $\psi_i(\tau,\sigma)=\phi_i(z)+\bar{\phi}_i(\zb)$
and find in accordance with (\ref{bc}) the periodicity properties
\begin{equation}
  \psi_i(\tau,\sigma+2\pi)=\psi_i(\tau,\sigma)+\frac{2\pi w}{\gamma}
  \delta_{i,2},
\end{equation}
which assign $\psi_1$ and $\psi_2$ directly to $r(z)$ respectively 
$t(z)$.The corresponding mode expansions are given by
\begin{eqnarray}
  \phi_i(z)&=&\frac{1}{2}q_i+(\frac{1}{4\pi}p_i+\frac{w}{2\gamma}
  \delta_{i,2})z+\frac{\i}{\sqrt{4\pi}}\sum_{n\neq0}
  \frac{{a_n}^{(i)}}{n}\e^{-\i nz}, \nonumber\\
 \bar{\phi}_i(\zb)&=&\frac{1}{2}q_i+(\frac{1}{4\pi}p_i-\frac{w}{2\gamma}
  \delta_{i,2})\zb+\frac{\i}{\sqrt{4\pi}}\sum_{n\neq0}
  \frac{{\bar{a}_n}^{(i)}}{n}\e^{-\i n\bar{z}}.
\end{eqnarray}
This tells us that $A(z)$ is periodic up to a factor (comp.\ (\ref{ABinphi}))
\begin{equation}
  A(z+2\pi)=\exp\left(-\gamma p_1\right)A(z).
\end{equation}
The integration of $A'(z)$ then yields $A(z)$ non-locally related to the
free fields
\begin{eqnarray}
  \lefteqn{A(z)=-\frac{\gamma}{2}\sh{-1}{\left(\frac{\gamma p_1}{2}
\right)}\times}  \nonumber\\ 
  &&\hspace{.2cm}\int_{0}^{2\pi}\!\!\!\d z'
\exp\left(-\frac{\gamma p_1}{2}\epsilon(z-z')\right)
\left(\delzp\phi_1-\i\delzp\phi_2\right)
  \exp\left(-2\gamma\phi_1(z')\right).
\end{eqnarray}
With
\begin{equation}
  \delz\exp\left(\alpha\;\epsilon(z-z')\right)=2\sh{}{\alpha}
  {\delta}_{2\pi}(z-z')\exp\left(\frac{\alpha}{\pi}\;(z-z')\right), 
\end{equation}
we get back the local $A'(z)$ of (\ref{ABinphi}). As seen from 
(\ref{nuint}) the conserved $V_\pm(z)$ are not periodic, except
the zero mode of the angle variable $\psi_2$ given by  
$\gamma p_2=\int_{0}^{2\pi}\!\!\!\d z'\dot{t}\tanh^2r$
becomes discrete $\gamma p_2=2\pi n$, $n\in{\mathbb{Z}}$. 

The general solution of the dynamical eqs.\ (\ref{eom}) is only 
complete after we have included the corresponding anti-chiral parts.
Choosing different target-space coordinates 
\begin{equation}
u=\sinh r\;\e^{\,\i t},\quad \bar{u}=\sinh r\;\e^{-\i t},
\end{equation}
the (covariantly written) action (\ref{action}) becomes
\begin{equation}
  S=\frac{1}{2\gamma^2}\int_{}^{}\d\tau\d\sigma\sqrt{-g}g^{ij}
  \frac{\partial_iu\;\partial_j\bar{u}}{1+u\bar{u}}.
\end{equation}
It shows for Wick rotated $t$ the metric singularity
of a black hole \cite{Witten}.
In terms of these new coordinates the solution (\ref{solution}, 
\ref{tosolution}) gets a rather simple form if it is expressed
by $y_1(z)$, $y_2(z)$ (\ref{Ainy}) (respectively $\bar{y}_1(\zb)$, 
$\bar{y}_2(\zb)$)
\begin{equation}
u=\frac{\bar{y}_1y_1'+\bar{y}_2y_2'}{y_1y_2'-y_1'y_2}, \qquad
\bar{u}=\frac{y_1\bar{y}_1'+y_2\bar{y}_2'}
{\bar{y}_1\bar{y}_2'-\bar{y}_1'\bar{y}_2}.
\end{equation}
The GL(2,$\cc$) invariance of the solutions now translates into
\begin{equation}\label{gl2c-psi}
\left(\begin{array}{c}y_1\\y_2\end{array}\right)\to
\left(\begin{array}{rc}
d&c\\-b&a\end{array}\right)
\left(\begin{array}{c}y_1\\y_2\end{array}\right),\quad
\left(\begin{array}{c}
\bar{y}_1\\\bar{y}_2\end{array}\right)\to
\left(\begin{array}{rc}
a&b\\-c&d\end{array}\right)
\left(\begin{array}{c}
\bar{y}_1\\\bar{y}_2\end{array}\right).
\end{equation}
Introducing the complex free fields
\begin{equation}
\psi=\psi_1+\i\psi_2,\qquad
\bar{\psi}=\psi_1-\i\psi_2,
\end{equation}
we find a B\"acklund transformation of the equations of motion of the 
$u$,$\bar{u}$
\begin{equation}\label{bl}\hspace{-0.5ex}
\begin{array}{ll}
\displaystyle
\delz u=\frac{\gamma}{2}\;
\e^{\gamma\psi}\left(P+\i\sqrt{4Q-P^2}\right)\delz\psi,&
\displaystyle
\hspace{0ex}\delzb u=\frac{\gamma}{2}\;
\e^{\gamma\psi}\left(P-\i\sqrt{4Q-P^2}\right)\delzb\psi,\nonumber\\[3ex]
\displaystyle
\delz\bar{u}=\frac{\gamma}{2}\; 
\e^{\gamma\bar{\psi}}\left(P-\i\sqrt{4Q-P^2}\right)\delz\bar{\psi},&
\displaystyle
\hspace{0ex}\delzb\bar{u}=\frac{\gamma}{2}\;
\e^{\gamma\bar{\psi}}\left(P+\i\sqrt{4Q-P^2}\right)\delzb\bar{\psi},
\end{array}\hspace{-2.5ex}
\end{equation}
where
\begin{equation}\label{pq}
P(u,\bar{u},\psi,\bar{\psi})=
u\,\e^{-\gamma\psi}+\bar{u}\,\e^{-\gamma\bar{\psi}}+
\e^{-\gamma\psi-\gamma\bar{\psi}},\quad
Q(u,\bar{u},\psi,\bar{\psi})=
(1+u\bar{u})\;\e^{-\gamma\psi-\gamma\bar{\psi}}.
\end{equation}
The integrability conditions for the fields $u$ and $\bar{u}$ are
equivalent to the free field equations for $\psi$ and $\bar{\psi}$, and
those for the $\psi$ and $\bar{\psi}$ provide the non-linear equations
of motion. The B\"acklund transformation (\ref{bl}, \ref{pq}) respects
periodicity.

Although the classical solution of the \slu{} gauged WZNW model
(\ref{action}) bears strong resemblance to Liouville or Toda theories 
its quantum structure might be different because its energy-momentum 
tensor does not have a `central charge' term classically. Furthermore,
we would like to see whether the black hole singularity survives 
quantization, a dilaton might appear, and whether the Hawking 
radiation could  be discussed in some manner based on this 
analytical solution.

\end{document}